\documentclass[aps,pra,showpacs,twocolumn,superscriptaddress,amsmath,amssymb, floatfix]{revtex4-1}
\usepackage{graphicx}
\usepackage{epstopdf}
\usepackage{subfigure}
\usepackage{dcolumn}
\usepackage{bm}
\usepackage{extarrows} 
\usepackage{hyperref}
\hypersetup{colorlinks=true,       
   linkcolor=blue,          
   citecolor=blue,        
   urlcolor=blue }	

\def\be{\begin{equation}}      
\def\ee{\end{equation}}
\def\bea{\begin{eqnarray}}      
\def\eea{\end{eqnarray}}

\def\beu{\begin{equation*}}   
\def\eeu{\end{equation*}}

\providecommand{\mr}[1]{\mathrm{#1}}
\providecommand{\abs}[1]{\left\lvert#1\right\rvert}   
\providecommand{\ket}[1]{\left|#1\right\rangle}

\providecommand{\bra}[1]{\left\langle#1\right|}
\providecommand{\mean}[1]{\left\langle#1\right\rangle}

\providecommand{\kB}{k_{\mathrm{B}}}

\renewcommand{\section}[1]{\textbf{#1}--}
\renewcommand{\subsection}{}

\begin{document}
\title{Cooling a harmonic oscillator by optomechanical modification of its bath}


\author{Xunnong Xu }
\affiliation{Joint Quantum Institute, University of Maryland/National Institute of Standards and Technology, College Park, Maryland 20742, USA}
\author{Thomas Purdy}
\affiliation{Joint Quantum Institute, University of Maryland/National Institute of Standards and Technology, College Park, Maryland 20742, USA}
\author{Jacob M. Taylor}
\affiliation{Joint Quantum Institute, University of Maryland/National Institute of Standards and Technology, College Park, Maryland 20742, USA}
\affiliation{Joint Center for Quantum Information and Computer Science, University of Maryland, College Park, Maryland 20742, USA}



\date{\today}
\begin{abstract}
Optomechanical systems show tremendous promise for high sensitivity sensing of forces and modification of mechanical properties via light.  For example, similar to neutral atoms and trapped ions, laser cooling of mechanical motion by radiation pressure can take single mechanical modes to their ground state. Conventional optomechanical cooling is able to introduce additional damping channel to mechanical motion, while keeping its thermal noise at the same level, and as a consequence, the effective temperature of the mechanical mode is lowered.  However, the ratio of temperature to quality factor remains roughly constant, preventing dramatic advances in quantum sensing using this approach.  Here we propose an approach for simultaneously reducing the thermal load on a mechanical resonator while improving its quality factor. In essence, we use the optical interaction to dynamically modify the dominant damping mechanism, providing an optomechanically-induced effect analogous to a phononic band gap. The mechanical mode of interest is assumed to be weakly coupled to its heat bath but strongly coupled to a second mechanical mode, which is cooled  by radiation pressure coupling to a red detuned cavity field. We also identify a realistic optomechanical design that has the potential to realize this novel cooling scheme.

\end{abstract}


\maketitle

Recent years have seen dramatic experimental and theoretical progress in optomechanics \cite{Kippenberg2008, Aspelmeyer2014}, ranging from ground state cooling \cite{Chan2011} and squeezing \cite{Purdy13,Safavi-Naneini2013} to quantum nonlinear optomechanics \cite{Nunnenkamp11, Rabl2011, Kronwald2013, Lemonde2013,Borkje2013, Xu2015}. These advances rely upon improvements in optomechanical coupling, particularly the single phonon-single photon coupling rate, and upon increasing mechanical quality factor, which enables lower heat loads and corresponds to higher sensitivity and longer quantum coherence times. However, the longer-term target of single photon nonlinear optics with optomechanical systems remains out of reach. Furthermore, for many sensing applications, the thermal noise remains a fundamental limit for relevant resonator designs, regardless of progress in the use of quantum correlations~\cite{Purdy13,Safavi-Naneini2013,Xu2014}, as typically the signal to be sensed is transduced to a force on the mechanical system which is in competition with the quantum Brownian motion-induced Langevin force from the thermal bath.

In the present work, we shall focus on thermal noise reduction for mechanical resonators, utilizing the standard tool box provided by optomechanics. This is crucial for improving the signal-to-noise ratio  of  mechanical devices, operating either in the classical regime or in the quantum regime. We are motivated by recent advances in phononic-band gap engineering as a principle for improved quality factor~\cite{Yu2014,Safavi-Naeini14,Gomis-Bresco14} -- but here, we engineer the band-gap via the optomechanical interaction, rather than during fabrication. Specifically, we introduce a generic coupled-oscillator model to describe mechanical systems whose damping is primarily via elastic wave radiation through the boundary, i.e., clamping loss. We then consider how optomechanical coupling to the clamping region enables dynamical control over the coupled mechanical resonator. This leads to the counterintuitive outcome: increasing optical power simultaneously reduces the temperature and linewidth of the mechanical mode, in contrast to direct optomechanical cooling. After introducing this model, we describe a specific resonator design that enables testing of these concepts using current techniques, and analyze the regime in which clamping losses are likely to dominate, finding that a low temperature and high mechanical frequencies our approach may find wide application.

\section{Toy model}
\begin{figure}[h]
\begin{center}
\includegraphics[width=0.95\columnwidth]{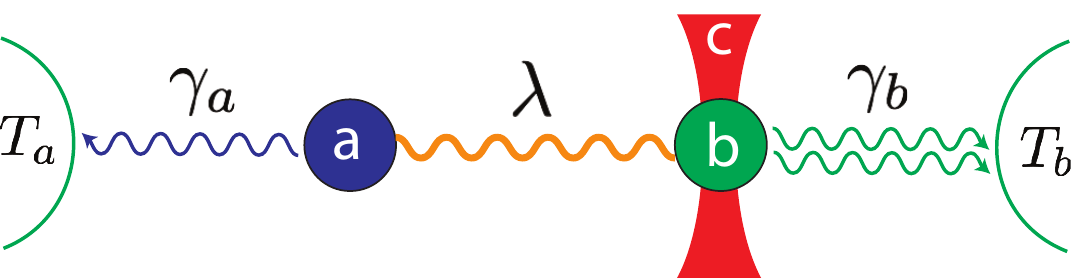}
\caption{(color online). Schematic of the coupled harmonic oscillator system, with optomechanical cooling on oscillator $b$. We are interested in the regime where $a$ couples weakly to its heat bath, which means $\gamma_a \ll \lambda, \gamma_b$. } 
\label{fig:fig1}
\end{center}
\end{figure}

We consider a toy model of two coupled quantum harmonic oscillators with annihilation operators $a$ and $b$, resonant frequencies $\omega_a$ and $\omega_b$, and a coupling strength between them $\lambda$. Each harmonic oscillator is also coupled to its own heat bath at temperature $T_a$ and $T_b$ with rates $\gamma_a$ and $\gamma_b$. In addition, optomechanical cooling is introduced to oscillator $b$ via coupling to a red detuned optical mode $c$ with frequency $\omega_c$ and damping $\kappa$,  as shown in Fig.~(\ref{fig:fig1}). 

The effective Hamiltonian of the three mode system when pumped with a laser follows immediately (with $\hbar=1$, and neglecting the weak nonlinear correction): 
\bea \label{hamiltonian}
H_{\text{eff}}  &=& (-\Delta  - i\frac{\kappa}{2}) c^{\dag} c + (\omega_a - i\frac{\gamma_a}{2}) a^{\dag} a + (\omega_b -i\frac{\gamma_b}{2})  b^{\dag} b  \nonumber \\
&& + \lambda (a + a^{\dag})(b + b^{\dag})  - \alpha g_0 (b+b^{\dag}) (c + c^{\dag}). 
\eea
where $\alpha = E/(i\Delta - \kappa/2)$ is the pump-induced coherent state in the optical cavity and assumed to be real without loss of generality (by choosing an appropriate phase for the pump strength $E$), $g_0$ is the quantum optomechanical coupling, and $\Delta = \omega_p - \omega_c$ is the detuning of the pump laser.
We consider $\omega_a \sim \omega_b \sim -\Delta$, so under the rotating wave approximation, the Heisenberg-Langevin equations in the input-output formalism are as follows: 
\begin{subequations} \label{Langevin}
\bea	
\dot{c} &=& i\Delta c - \frac{\kappa}{2} c + i \alpha g_0  b + \sqrt{\kappa} c_{\mathrm{in}} ,\\
\dot{a} &=& -i\omega_a a - \frac{\gamma_a}{2} a - i\lambda b + \sqrt{\gamma_a} a_{\mathrm{in}} ,\\
\dot{b} &=& -i\omega_b b - \frac{\gamma_b}{2} b - i\lambda a + i\alpha g_0 c +  \sqrt{\gamma_b} b_{\mathrm{in}} . 
\eea
\end{subequations}
This set of linear equations can be solved by moving to the frequency domain. We successively solve for $c(\omega)$, then $b$, then $a$. For example, $c(\omega) = \frac{i \alpha g_0 b +  \sqrt{\kappa} c_{\text{in}}}{-i(\omega+ \Delta) + \kappa/2}$. We immediately set $c \approx 
\frac{1}{\alpha g_0} (i \frac{\Gamma}{2} b + \sqrt{\Gamma} c_{\text{in}})$ in the sideband-resolved limit with $|\Delta + \omega| \ll \kappa/2$ where $\Gamma = 4 |\alpha g_0|^2/\kappa$ is the optically-induced damping of mode $b$. Continuing, we find
\begin{subequations}
\begin{align}
\chi_b^{-1} b(\omega) &= \sqrt{\gamma_b} b_{\text{in}}  + i\sqrt{\Gamma} c_{\text{in}}  \\
\text{where}\ \chi_b &= \left[ -i (\omega - \omega_b) + (\gamma_b + \Gamma)/2 \right]^{-1}
\end{align}
\end{subequations}
is the susceptibility of mode $b$ for $\lambda = 0$.

Finally, we find for mode $a$
\begin{subequations}
\begin{align} \label{solution} 
\chi_a^{-1} a &=  \sqrt{\gamma_a} a_{\text{in}}  - i \chi_b \lambda \left( \sqrt{\gamma_b} b_{\text{in}} + i\sqrt{\Gamma} c_{\text{in}} \right)\\
\text{with}\ \chi_a &= \left[ -i (\omega - \omega_a) + \gamma_a/2 + \chi_b \lambda^2 \right]^{-1}
\end{align}
\end{subequations}
Examining these equations, we see that mode $a$'s resonant response, as described by the susceptibility $\chi_a$, have a frequency and damping that depend, via $\lambda^2 \chi_b$, upon the properties of the optomechanically damped mode $b$. Specifically, examining the real and imaginary components, we have  
\begin{subequations}
\bea
\omega_a^{\prime} & =& \omega_a  + \frac{\lambda^2 (\omega - \omega_b) }{(\omega - \omega_b )^2 +  (\gamma_b + \Gamma)^2/4},  \\
\gamma_a^{\prime} & = & \gamma_a +  \frac{\lambda^2 (\gamma_b + \Gamma) }{(\omega - \omega_b )^2 +  (\gamma_b + \Gamma)^2/4},
\eea
\end{subequations}

Let us examine the particular scenario when the cooperativity between $a$ and $b$ satisfies $\mathcal{C}_{ab} \equiv \frac{4 \lambda^2}{\gamma_a\gamma_b} \gg 1$ and $\frac{\gamma_b + \Gamma}{\gamma_a} \gg \mathcal{C}_{ab} $. This corresponds to the intrinsic damping of mode $a$ being dominated by its coupling through $b$ to $b$'s bath, while simultaneously being able to examine $b$'s response as broader than $a$'s. We will further focus on $|\omega_b - \omega_a| \ll \gamma_b + \Gamma$, as provides the maximum modification of damping.  This allows us to expand $\omega_a' \approx \omega_a$ and $\gamma_a' \approx \gamma_a + \Gamma_a$ with $\Gamma_a \equiv \frac{4 \lambda^2}{\gamma_b + \Gamma}$. We finally get
\be  \label{solution}
a \approx \frac{\sqrt{\gamma_a} a_{\text{in}} + i \sqrt{\Gamma_a \left(\frac{\gamma_b}{\gamma_b + \Gamma}\right)} b_{\text{in}} +  \sqrt{\Gamma_a \left(\frac{\Gamma}{\gamma_b + \Gamma}\right)} c_{\text{in}} }{-i(\omega- \omega_a) + (\gamma_a + \Gamma_a)/2}
\ee

This regime (damping of $a$ primarily via mode $b$, which in turn is damped optically by a sideband-resolved coupling to mode $c$) lets us examine the effective temperature. Specifically, using the input noise correlations of $b_\mathrm{in}$ in the frequency domain, 
\begin{subequations}  \label{correlation} 
\bea  
\mean{b_\mathrm{in}^{\dag} (\omega)b_\mathrm{in}(\omega^{\prime})} &=& \bar{n} \delta(\omega+\omega^{\prime}) \\
\mean{b_\mathrm{in}(\omega)b_\mathrm{in}^{\dag} (\omega^{\prime})} &=& (\bar{n}+1) \delta(\omega+\omega^{\prime})
\eea
\end{subequations}
where $\bar{n} = 1/(e^{\hbar\omega/\kB T}-1)$ is the average phonon occupation number of a harmonic oscillator of frequency $\omega$ when it is in thermal equilibrium with a heat bath at temperature $T$. We have the same for $a_{\text{in}}$ and for $c_{\text{in}}$, though we assume $T=0$ for the last, while $T_a = T_b = T$ for the first two.
Now we find the average position fluctuation $n_{\text{eff}} + 1/2 \equiv \mean{(a+ a^\dag)^2}/2$ (since $\bra{n} (a + a^{\dag} )^2\ket{n}= 2n+1$) to be
\begin{equation}
n_{\text{eff}}  = \frac{
\gamma_a \bar{n}  + \Gamma_a \left( \frac{\gamma_b}{\Gamma+\gamma_b} \bar{n}  \right)
}{\gamma_a + \Gamma_a}
\end{equation}
where $\bar{n}$ is evaluated at $\omega_a$. This expression is also consistent with the result from detailed balance relation \cite{Marquardt2007}.  We can minimize this occupation by a setting the optomechanical  cooperativity $\mathcal{C}_{\mathrm{OM}} \equiv \Gamma/\gamma_b$ of $b$ to  
\[
\mathcal{C}_{\mathrm{OM}}  \rightarrow \mathcal{C}_{\mathrm{OM}} ^* \equiv \sqrt{ 1 +\mathcal{C}_{ab}  }\ ,
\]
which gives
\[
\frac{n_{\text{eff}}^* }{\bar{n}} = \frac{2}{1 + \sqrt{1 +\mathcal{C}_{ab}}}. 
\]
Thus in principle even ground state cooling is achievable, if the mechanical oscillator cooperativity $ \mathcal{C}_{ab} \gtrsim 16 \bar{n}^2$. Curiously, this cooling arises with \emph{reduction} of the linewidth of mode $a$, with the limiting linewidth $\gamma_a + \Gamma_a^* = \gamma_a \sqrt{ 1 + \mathcal{C}_{ab}}$.

Finally, to confirm these approximations, we numerically examine the same regime, but without making the rotating wave approximation or any narrowband approximations -- this enables us to include counterrotating terms and their associated heating. We plot the rescaled position fluctuation spectrum $S_{xx}(\omega) = \int_{-\infty}^{+\infty} \text{d}t e^{i\omega t} \mean{x(t) x(0)} $ and rescaled effective temperature $T_{\text{eff}}/T$  below in Fig.~(\ref{fig:fig2}). We find that when $g_0 \alpha, \lambda \ll \omega_a, \omega_b$ our approximate theory and the exact results are in agreement. 
\begin{figure}[!h]
\begin{center}
\includegraphics[width=0.98\columnwidth]{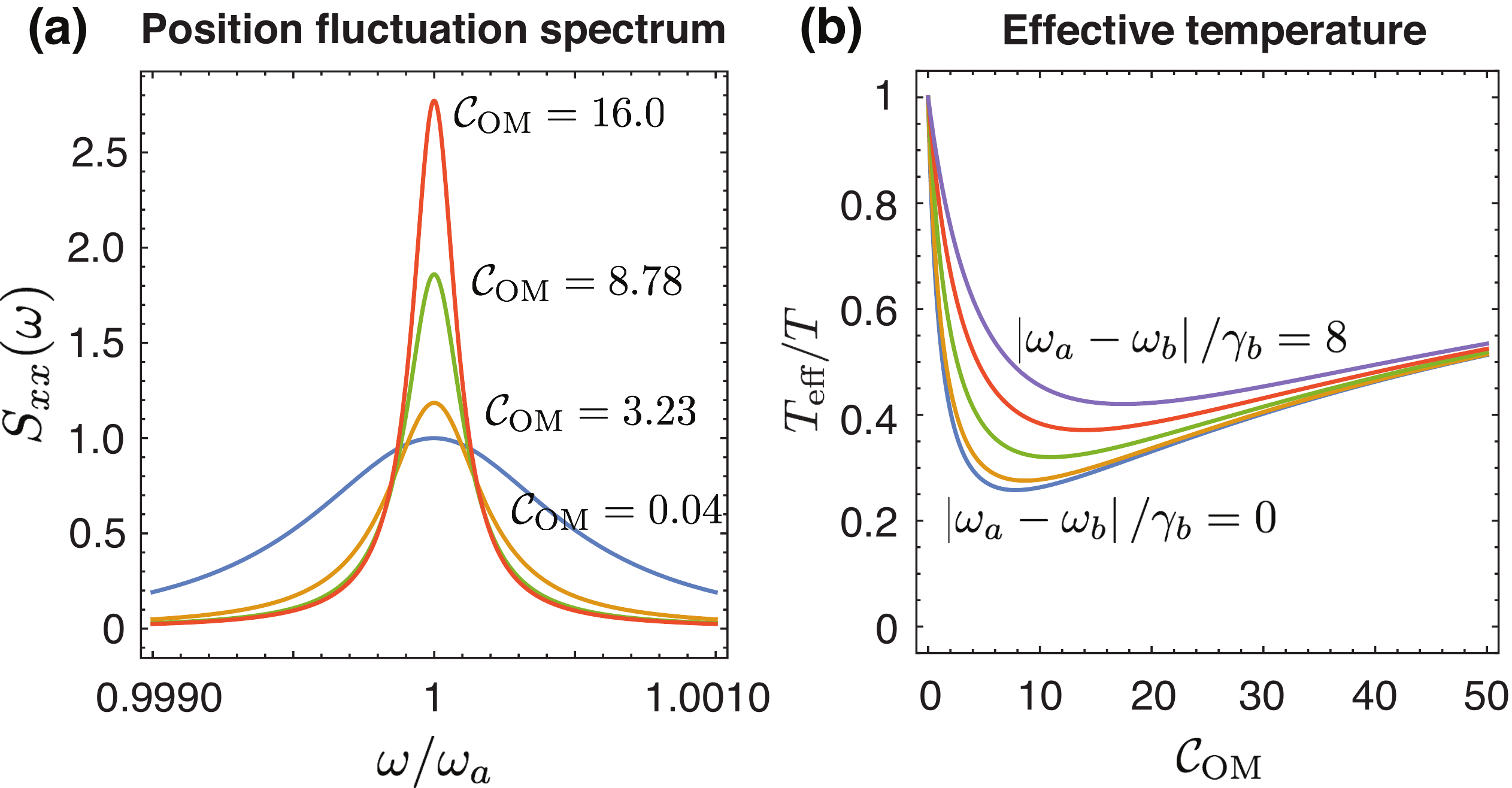}
\caption{(color online). (a) Rescaled position fluctuation spectrum for different values of optomechanical cooperativity $\mathcal{C}_{\mathrm{OM}}$, with the mechanical oscillator cooperativity chosen as $\mathcal{C}_{\mathrm{ab}} = 8$; (b) Rescaled effective temperature as a function of  $\mathcal{C}_{\mathrm{OM}}$ for $\mathcal{C}_{\mathrm{ab}} = 50$ and different values of  $\abs{(\omega_a - \omega_b)}/\gamma_b$.}
\label{fig:fig2}
\end{center}
\end{figure}

\section{Example implementation}
\begin{figure}[htp!]
\begin{center}
\includegraphics[width=0.98\columnwidth]{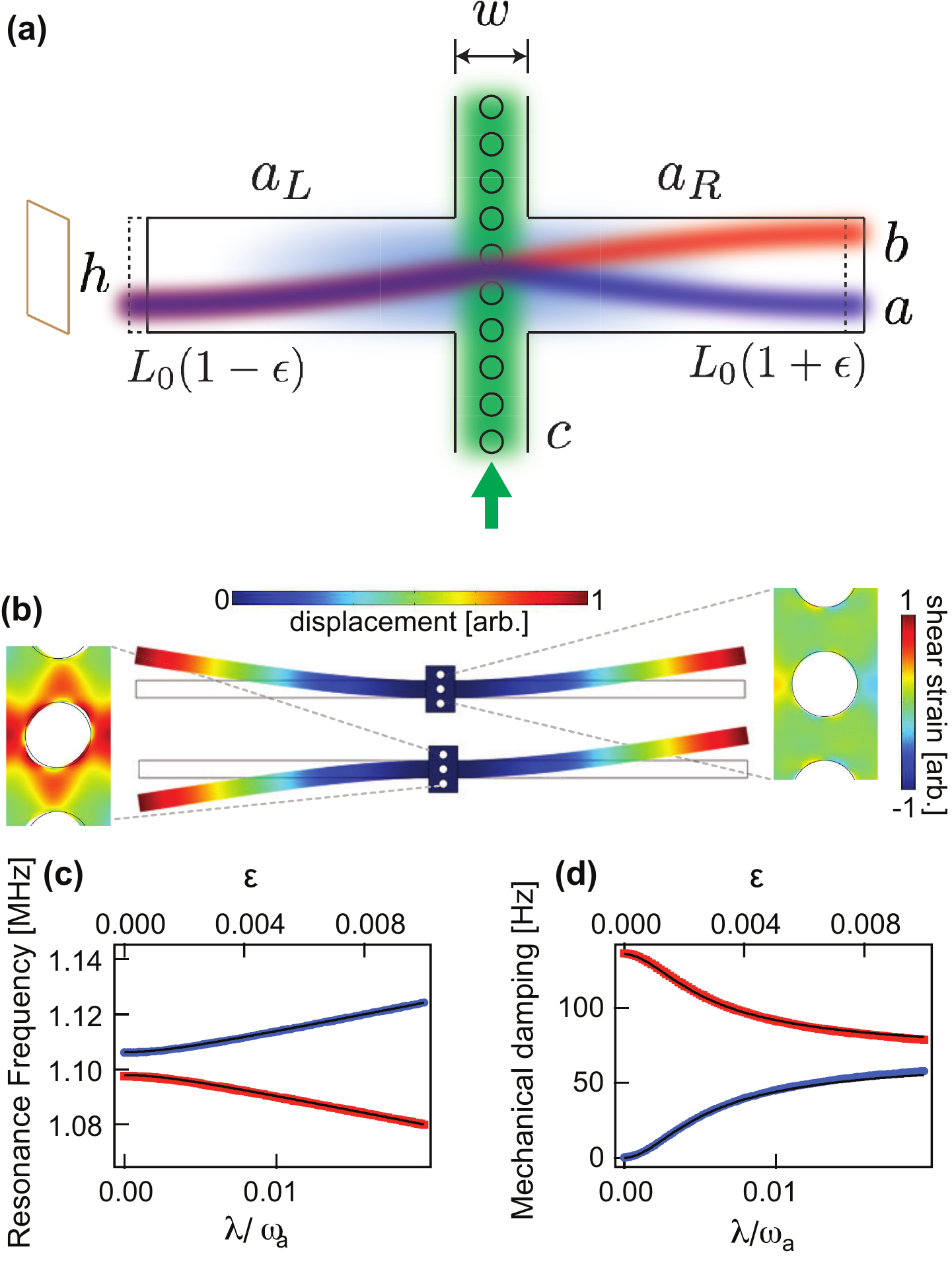}
\caption{(color online). (a) Optomechanical design consisting of two similar quarter-wave beam resonators, nominal length $L_0$, coupled through a center support region, width $w$, containing a photonic crystal optical structure. (b) Simulated symmetric (top) and antisymmetric (bottom) eigenmodes, $\epsilon=0$. Insets show the strain deformation of a single photonic defect (envisioned as part of a photonic crystal resonator) that would lead to a strong strain-induced optomechanical coupling only for the antisymmetric mode. (c,d) As the asymmetry, $\epsilon$, is increased, the antisymmetric and symmetric modes are increasingly coupled, shifting the simulated eigenfrequencies and clamping losses (red squares, blue circles), consistent with fits to the theoretical model of Eq.~(\ref{hamiltonian}) (black). Simulation parameters are $L_0$=20~$\mu$m, $h$=0.3~$\mu$m, and $w=0.5$~$\mu$m, corresponding to $\omega_0=2 \pi \times 1.102$~MHz and $\gamma_b/2=2 \pi \times 70$~Hz clamping loss for an individual arm fabricated from silicon nitride.
}
\label{fig:fig3}
\end{center}
\end{figure}

To design an optomechanical system that captures the main features of the toy model, we need three basic components: i) two coupled mechanical resonators; ii) one resonator is limited by thermoelastic damping, and the other is limited by clamping loss; iii) optomechanical cooling primarily on the second resonator. With these goals in mind, we can design a system that is shown in Fig.~(\ref{fig:fig3}). In this design, there are two nearly identical quarter-wave mechanical resonators on the left arm and right arm of a large beam resonator, denoted as $a_L$ and $a_R$. The lengths of the two arms may not be exactly the same (or the left-right symmetry may be broken by defects). This asymmetry leads to different resonant frequencies for the two resonators, with  $\omega_L = \omega_0(1+\epsilon)$ and $\omega_R=\omega_0(1-\epsilon)$, as shown in Fig.~(\ref{fig:fig3}). 

To examine the mode structure, we consider coupling between the left and right sides through the support structure with a strength $J$.
We see that symmetrical coupling of $a_L$ and $a_R$ through the support leads to normal modes $a = 1/\sqrt{2} (a_L + a_R)$ and $b = 1/\sqrt{2} (a_L - a_R)$, with the former having no clamping loss for $\epsilon = 0$ and the latter a clamping loss $\gamma_b \approx J^2 \rho$ (by Fermi's golden rule). Physically, the symmetric mode has destructive interference which prevents excitation of support structure and the associated clamping loss, analogous to the reduction in damping observed in tuning-fork resonators, which can be seen in Fig.~(\ref{fig:fig3}d) below. 

Meanwhile, the asymmetry couples $a$ and $b$ together with a rate $\lambda = \epsilon \omega_0$. In addition, there can be some intrinsic damping of mode $a$ via, e.g., thermoelastic loss with a rate $\gamma_a$, while we account for most of the damping of $b$ via clamping loss to the quasi-mode with rate $\gamma_b$. Under the rotating wave approximation, the mechanical parts of the Hamiltonian are the same as those in Eq.~(\ref{hamiltonian}).  

In our proposed structure, the optical mode $c$ is, e.g., a photonic crystal cavity in the center of the beam (the support structure) and is coupled strongly to the antisymmetric mode, as shown in the simulation below in Fig.~(\ref{fig:fig3}b).
We notice that the coupling rate $\lambda$ is proportional to the dimensionless asymmetry $\epsilon = \abs{L_L - L_R}/(L_L + L_R)$ in the length of the two arms of the beam resonator. In principal, $\lambda$ could be engineered in a wide range by fabricating resonators of different geometries or by temperature tuning of the length and speed of sound in the material.

As the toy model suggests, the damping of oscillator $a$ is assumed very weak ($\gamma_a\ll \gamma_b, \lambda$), so that the optomechanical cooling of oscillator $b$ could be effectively ``exported'' to $a$ through phonon tunneling. This naturally leads us to the question: what kind of oscillator design has this property? There are two main sources of mechanical damping in micro- and nano-mechanical resonators~\cite{Joshi14}: i) Boundary damping, or clamping loss, e.g elastic wave radiation from the material to its base through the boundary, and ii) material damping, which includes thermoelastic damping (TED), phonon-phonon interactions. The clamping loss represents the coupling from a resonator to its base, since phonons are exchanged through the boundary, while thermoelastic damping is the major contribution to the internal damping rate of a resonator. 

Clamping loss has been studied extensively in the literature \cite{Yasumura00, Hao2003, Chandorkar08, Ko11}. For a beam resonator where the thickness of the beam resonator is much smaller than the wavelength of the elastic wave propagating in its support,  the flexural vibration can be described using the ideal beam theory. The support of clamping-free (C-F) beam resonators is usually modeled as semi-infinite and infinite thin-plate, respectively, with the same thickness as the beam resonator; all the vibration energy of a beam resonator entering the support structure is considered to be lost. It is the vibrating shear force that induces this energy loss. In \cite{Hao2003}, they studied the clamping loss using elastic wave radiation theory and found the quality factor of  clamping-free (C-F) beam resonators to be: 
\be
Q_{\mathrm{C-F}} \propto \left( L/h\right)^2
\ee
where $L$ is the length of the beam and $h$ is its width.  

Secondly, we look at the thermoelastic damping \cite{Zener38, Srikar02}. Phonons traveling through a large elastic material will experience damping due to their nonlinear interaction with a surrounding bath of phonons. In the ‘‘diffusive’’ regime where the mean free path of these thermal phonons is much smaller than the wavelength of the acoustic mode, the interaction between the phonon mode and the thermal bath is captured by the material's thermal expansion coefficient (TEC), defined as $\alpha \equiv  \frac{1}{L} \frac{\partial L}{\partial T}$, which is temperature dependent. According to \cite{Lifshitz2000}, the quality factor corresponding to this damping mechanism is given by
\be \label{thermal_damping}
Q_{\mathrm{TED}}^{-1} = \frac{E\alpha^2 T}{C_p} f(h/h_0), 
\ee
where $E$ is the material's Young's modulus, $T$ is the temperature, $C_p$ is the  heat capacity at constant pressure, and $f(h/h_0)$ is a beam geometry function parametrized by a critical beam width $h_0$. 

A detailed numerical estimate of these losses for a specific mechanical resonator such as SiN is possible, but here we remark that for short beams, the clamping loss, which grows as $h^2/L^2$, will always tend to dominate over the thermoelastic damping. For example, for a resonator with frequency $\Omega/2\pi = 1~\mathrm{MHz}$, we have $h_0 = 6.546~mm$. When $h\ll h_0$, we find $f(h/h_0) \to 5h^2/h_0^2$, which gives us a very high $Q_\mathrm{TED}$ (well beyond the usual material limits). 

\section{Analysis of force sensing}
The proposed scheme for reducing the thermal load of the mechanical oscillator is useful for force sensing, where thermal noise is a main obstacle towards building ultra-sensitive force detection devices. Specifically, we now show that we can achieve a lower thermal noise floor than is possible with conventional optomechanical cooling. 

There exist a simple relation between the ultimate sensitivity (in units of $N/\sqrt{\mathrm{Hz}}$) for a mechanics based device and its thermal noise level, which can be calculated as the power spectral density of thermal fluctuating forces:
\be
s(\omega) \equiv \sqrt{S_{{FF}}(\omega)} =  \sqrt{\int_{-\infty}^{+\infty} \text{d}t e^{i\omega t} \mean{F(t) F(0)} } 
\ee
In the case of our coupled harmonic oscillator system, if oscillator $a$ is used for force sensing, then we get better sensitivity because of the reduction in its thermal load. The force on a harmonic oscillator is defined as 
\be
F = \dot{p} =  -i \sqrt{\frac{\hbar m\omega_a}{2}} (\dot{a} - \dot{a}^{\dag}), 
\ee
so the corresponding fluctuating force in frequency domain can be found from Eq.~(\ref{solution}) 
\bea
F_{\text{in}}(\omega) &=&  -i \sqrt{\frac{\hbar m\omega_a}{2}} \left[\sqrt{\gamma_a} a_{\rm{in}} + i\sqrt{\Gamma_a\left(\frac{\gamma_b}{\gamma_b + \Gamma} \right)} b_{\rm{in}} - \mathrm{h.c.}  \right]  \nonumber \\
\eea
Using the noise correlation functions Eq.~(\ref{correlation}), we can calculate its spectral density in the narrow band limit as 
\bea
S_{FF} (\omega) &=& \int \mr{d} \omega^{\prime}  \mean{\abs{F_{\text{in}}(\omega) F_{\text{in}}(\omega^{\prime}) } }\nonumber \\
&=& \frac{\hbar m\omega_a}{2}\left[ \gamma_a   + \Gamma_a \left(\frac{\gamma_b}{\gamma_b + \Gamma} \right) \right]  (2\bar{n} + 1) \nonumber \\
&\approx & m \left[ \gamma_a + \frac{4\lambda^2\gamma_b}{(\gamma_b + \Gamma)^2} \right] k_{\rm{B}} T \nonumber \\
&=&  \left[ 1 + \frac{\mathcal{C}_{ab}}{(1 + \mathcal{C}_{\mathrm{OM}})^2} \right] m\gamma_a k_{\rm{B}} T \label{e:noise}
\eea
where we recall $\mathcal{C}_{\mathrm{OM}} = \Gamma/\gamma_b$ is the optomechanical cooperativity and  $\mathcal{C}_{ab} = \frac{4\lambda^2}{\gamma_a\gamma_b}$ is the cooperativity between $a$ and $b$. When the optically induced damping rate $\Gamma$ is large compared to $\gamma_b$, we have substantial noise reduction and thus improved sensitivity for the device compared to conventional optomechanical cooling. In the latter case, we could have the noise floor of Eq.~(\ref{e:noise}) but with $\mathcal{C}_{\mathrm{OM}} = 0$.

Here we proposed an efficient scheme for cooling a harmonic oscillator by decreasing dissipation via optomechanical cooling. We studied the practical conditions to realize this cooling scheme, and also identified a realistic optomechanical design that has the potential to realize it. Potential applications include mechanics based force sensing, and other related areas where reducing the thermal load via non-conventional techniques is need.

\section{Acknowledgements} 
We thank Albert Schliesser, Eugene Polzik and Kartik Srinivasan, John Lawall and Anders Sorensen for helpful discussions. Funding is provided by DARPA QuASAR and the NSF Physics Frontier at the JQI.

\bibliography{2HO_damping_v11B}

\end{document}